\documentclass[a4paper, 11pt]{article}
 \pdfoutput=1 
\usepackage{url}
\usepackage{tabularx}
\usepackage{booktabs}
\usepackage{graphicx}
\usepackage{color}
\usepackage{epsfig}
\usepackage{amssymb}
\usepackage{amsmath}
\usepackage{amsthm}
\usepackage{latexsym}
\usepackage{setspace}
\usepackage{bbm}
\usepackage[margin=1in,footskip=0.25in]{geometry}
\usepackage{float}
\usepackage{dsfont}
\usepackage{caption}
\usepackage{subcaption}
\usepackage{enumerate}
\usepackage[ruled,vlined]{algorithm2e}
\usepackage{hyperref}

\newtheorem*{corollary*}{Corollary}

\newtheorem*{lemma*}{Lemmas}

\theoremstyle{remark}
\newtheorem*{remark*}{Remark}
\newtheorem*{remarks*}{Remarks}
\theoremstyle{definition}

\newtheorem{remark}{Remark}

\allowdisplaybreaks

\title{How Reliable are Test Numbers for Revealing the COVID-19 Ground Truth and Applying Interventions?}

\author{Aditya Gopalan \and Himanshu Tyagi}
 \date{Indian Institute of Science,\\
 Bengaluru 560012, India\\
 		 \{aditya, htyagi\}@iisc.ac.in}

\begin{document}

\maketitle

 \begin{abstract}
   The number of confirmed cases of COVID-19 is often used as a proxy for the actual number of ground truth COVID-19 infected cases
   in both public discourse and policy making. However, the number of confirmed cases depends on the testing policy, and it is important
   to understand how the number of positive cases obtained using different testing policies reveals the unknown ground truth. 
   We develop an agent-based simulation framework in Python that can simulate various testing policies as well as interventions such as lockdown based on them. The interaction between the agents can take into account various communities and mobility patterns.
   A distinguishing feature of our framework is the presence of another `flu'-like illness with symptoms similar to COVID-19, that allows us to model the noise in selecting the pool of patients to be tested.  We instantiate our model for the city of Bengaluru in India, using census data to distribute agents geographically, and traffic flow mobility data to model long-distance interactions and mixing. We use the simulation framework to compare the performance of three testing policies: Random Symptomatic Testing (RST), Contact Tracing (CT), and a new Location Based Testing policy (LBT). We observe that if a sufficient fraction of symptomatic patients come out for testing, then RST can capture the ground truth quite closely even with very few daily tests. However, CT consistently captures more positive cases. Interestingly, our new LBT, which is operationally less intensive than CT, gives performance that is comparable with CT.
In another direction, we compare the efficacy of these three testing policies in enabling lockdown, and observe that CT flattens the ground truth curve maximally, followed closely by LBT, and significantly better than RST.
   \end{abstract}

\section{Introduction and Summary of Observations}
  Beginning March 11, fearing the quick spread of COVID-19, all 
schools were shut down in Bengaluru, India. Colleges, universities,
and cinema halls were soon to follow and were shut down within a
week. On Sunday March 22, the Prime Minister of India announced a
country-wide ``Janta curfew'' for a day. Finally, on Tuesday, March
24, the Prime Minister announced a 21 day complete lockdown for the
country, which has been extended till May 3rd now. It is interesting
to note that there were just 6 confirmed cases and 1 death till March 15 across the state of
Karnataka, to which Bengaluru belongs, when the mega-city of Bengaluru
was shut down. In fact, till March 24 there were only 517 confirmed
cases across India\footnote{These numbers have been taken
  from~\cite{covid19India}.}, a number which may appear small compared
to the large ($\approx$ 1.3 Billion) population of the country. Even the growth
rate of the number of cases was not very high.  Yet, the policy makers
decided to engage a lockdown, perhaps even with popular public
support.

Clearly, the perceived number of ground truth cases and its increase
rate must have been higher than the actual confirmed cases to
facilitate such a drastic measure.  But how well does the daily
positive test outcome trend reflect the unknown ground truth? Can 
policy makers use it reliably to implement non-pharmaceutical interventions,
such as lockdown, even when the number of daily tests is very low?
Such questions become even more important in view of several media
articles and expert opinions questioning if India is testing its residents 
enough \cite{nottesting1,nottesting2}. Given the large population of India, it is unlikely that we
can scale the tests to the level of countries like Iceland, which has
tested about 5\% of its population. Thus, what one expects out of more
testing is to get a better estimate of the ground truth trend to
enable better preparedness and well-informed policy decision
making. With this context in mind, we ask a basic question: How much testing is enough?

We undertake a systematic mathematical modeling and simulation-based
study of a variety of testing policies, and compare their efficacy for enabling
interventions such as lockdown. Based on our experiments, our main
findings can be summarized as follows:
\begin{enumerate}
\item If a sufficient fraction of symptomatic population shows up for
  testing, then testing a small random sample of symptomatic patients
  can give a good idea of the ground truth trend.

\item Contact tracing (CT), where contacts of a  COVID-19 patient are
  tested, returns a significantly higher number of positive test outcomes than the
  random symptomatic testing (RST) as above. More importantly, a
  decision for locking down the population based on CT can help reduce the peak ground
  truth number of cases (`flatten the curve') much better than RST.

\item By using a location- and mobility pattern-aware sampling, it is possible to get performance similar to that of CT using
  operationally less intensive testing procedures.
\end{enumerate}
We present the precise observations and describe the setup later in
the paper, but we quickly note here that we use the derivative of the
ground truth curve as an indicator of the ground truth trend --
indeed, visually it appears to be a good indicator. It is this
indicator of ground truth trend that RST reveals well.

Our conclusions are based on agent-based simulation for a population of 100000 individuals or {\em agents} distributed across a
realistic synthetic city, interacting based on a realistic mobility
pattern. Specifically, we use publicly available census data to distribute the agents
across the 198 (urban) wards in the city of Bengaluru. The agents have health 
states related to COVID-19 and another, generic, `flu'-like disease condition  with similar symptoms,
which evolve independently.  A susceptible agent (in COVID state S)
can get infect when it meets a COVID-19 infected agent (in COVID state
I).  The agents can interact with other agents in their neighbourhood
or agents that visit similar locations daily. We instantiate the
mobility of agents across the city using mobility data obtained from
traffic flows.

This evolution model drives the unknown state, a part of which is
observed by testing policies, stored in a separate module in our
implementation.  A testing policy determines which agents will be
subjected to testing and applies a randomized test to the selected
agent. The history of test results is stored and is made available to  intervention policies, which are stored in yet another module. An
intervention policy outputs a control action which modifies the state
evolution dynamics. For instance, a lockdown intervention will disable
the interaction between the agents. We assume that the borders of the city are closed
and there is no interaction with the outside world. 

Our overall simulation framework, made available as a Python package at~\cite{CovidSimRepo}, is flexible
and can incorporate any new testing policy or intervention policy. Furthermore, the state evolution model
can be easily modified to incorporate more ``mixing points'' such as buses, malls, {\it etc.}. 
In fact, our goal is to incorporate real-time mobility data obtained from digital platforms, as suggested in~\cite{Ferrettieabb6936}, to have an updated representation of mixing in the city.

A disclaimer: Our model has not been calibrated to match the actual
number of cases in Bengaluru. Neither do our conclusions enjoy
rigorous theoretical backing. These results are preliminary and are
based on experiments with our simulation framework. They provide, we hope, 
insight into questions raised above and exhibit the utility of our
simulation framework that can used for such studies. This is work in
progress, released early to ensure timely dissemination.

There is large body of work using mathematical models and simulation
to study spread of COVID-19 and facilitate decision making. A very
timely publication was the report~\cite{ferguson2020npis} from the COVID-19 response team of
Imperial College, London, earlier versions of
which raised an alarm about possible worst-case scenarios if COVID-19
is allowed to grow unchecked. This in turn is based on a long-line of
work from the same group on mathematical modeling and simulation of
epidemic spread; see, for instance,~\cite{ferguson2006influenza}. Our models of state
evolution are similar to these works, but at this point are not as
elaborate as these works. More refined models with comorbidity and
age-dependent evolution have been considered
in~\cite{singh2020agestructured}.

While distance based modeling of interaction of agents is quite
popular, with footing in random graph theory, 
using real mobility and traffic data to model interaction of agents is also gaining
prominence in epidemiological studies.
In this paper, we have only considered traffic data
obtained using surveys, somewhat similar to how data is obtained
in~\cite{Klepac2020.02.16.20023754}. A more effective method can be
the prescription in~\cite{Ferrettieabb6936} where location services and mobile
phone usage data is used to obtain real-time daily mobility
patterns. Looking ahead, we would like to integrate such data into our framework.

While most of prior work has treated the outcome of tests as the
actual ground truth number of COVID-19 cases, very recently, articles~\cite{lorch2020spatiotemporal}
and~\cite{deshpande2020} have appeared that explicitly study the role of the testing
policy. The former has a similar setup as ours, except that the
simulation is event-driven and uses a more detailed simulation of
agent interaction. However,  only 
contact tracing is considered and an ideal situation where there is no
other flu with similar symptoms is assumed. In a way, this framework
assumes that one can directly access the ground truth. The
article~\cite{deshpande2020} presents a similar discussion as our
results, but the approach appears to be more statistical and not based on explicit epidemic modeling/simulation. We remark that our proposed new testing algorithm uses
ideas from multi-armed bandit problems; the
paper~\cite{biswas2020covid19} proposes an algorithm that uses similar
ideas towards achieving a good test allocation strategy.

The remainder of this paper is organized as follows. We present the
details of our modeling and simulation framework in the next
section. A comparison of the three testing policies we consider when
there is no intervention is given in Section~\ref{s:testing}, and a
comparison of their efficacy in enabling interventions in
Section~\ref{s:interventions}. We conclude with some discussion on
policy implications and next steps in the final section.

\section{Simulator Description and Features}
\label{s:simulation}
\subsection{Simulation model}
We consider an agent based simulation framework to model the
propagation of an epidemic in a city. In our general framework, a city
is populated using $n$ agents distributed across fixed ``localities''
of a city in proportion to their population densities. Each agent $i$
represents a person
with attributes such as location associated with it. The health of an 
agent is captured by its ``COVID State'' $C_i(t)$ on day $t$. A
distinguishing feature of  
our setup is an additional ``Flu State'' $F_i(t)$ which represents the
presence 
of another flu with similar symptoms as COVID-19. This allows us to
model the erroneous prescription of COVID-19 test to a person who
shows similar symptoms due to presence of another flu, which we believe is 
essential for a realistic modeling of testing.

\begin{figure}[h]
  \centering
  \includegraphics[scale=0.8]{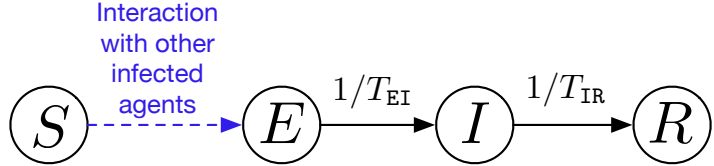}
 \caption{COVID state evolution of each agent.}
 \label{f:TPM-COVID}
\end{figure}

The evolution of $C_i(t)$ is determined by two factors: a local
evolution model for each agent's state and evolution due to
interaction between agents. 
For local evolution, 
we use the popular SEIR
model for $C_i(t)$ 
where the COVID state of an agent can take values in
the set $\{S,E,I, R\}$ representing, respectively, susceptible (S),
exposed (E), infected (I), and recovered (R) conditions.
Note that this model is a simplification of the dynamics used in~\cite{ferguson2020npis}
and combines the several stages such as hospitalization and death into a single state R.
The origin of such models in epidemiology can be traced back to the pioneering work of
Kermack and McKendrick in 1920s,
and they have been used for modeling COVID-19 dynamics
in~\cite{Ferrettieabb6936, Lieabb3221, singh2020agestructured, lorch2020spatiotemporal} 
The
collective states $C(t)=\{C_i(t), 1\leq i \leq n\}$, $1\leq t\leq T$, form a
discrete-time Markov chain where 
$C_i(t)$ on day $t$  changes to the state $C_i(t+1)$ on day
$t+1$ according to a pre-specified  probability transition matrix $T$
depicted in Figure~\ref{f:TPM-COVID}. 
Specifically, each agent changes its
state from E to I to R independently of all the other agents. The
transition probabilities in Figure~\ref{f:TPM-COVID} are set to parameters $1/T_{ss'}$ where $T$ is the
average transition time between states $s$ and $s'$. Although the values chosen for our simulations are only approximations, such average times have
been studied and reported extensively in literature, and the most up-to-date estimates can be used in the model by the interested practitioner.

Note that an agent makes a transition from S to E based
on its interaction with other agents. When an agent in state $I$ meets
another agent in state $S$, the latter agent gets infected with a
pre-specified probability $p$. To model the meeting of agents, we
include two components:  The first is a ``neighbourhood'' component where
each agent meets a set of randomly chosen agents from the same
neighbourhood. In addition, each agent meets a fixed set of agents from
its neighbourhood, generated randomly at start and then fixed throughout the
simulation. Note that, for an agent, its neighbourhood need not coincide with its locality
and can include a set of close-by localities. In our simulation, we
have defined a neighbourhood as a set of localities touching
(geographically) a given locality.

The second component represents interaction with agents
from different localities, not necessarily the neighbouring ones. Here
we propose to use data about mobility in the  city. Here, too, an agent
visiting a location on a day interacts with a set of randomly
chosen agents generated afresh everyday as well as fixed set of agents
set upfront.

\begin{figure}[h]
  \centering
  \includegraphics[scale=0.6]{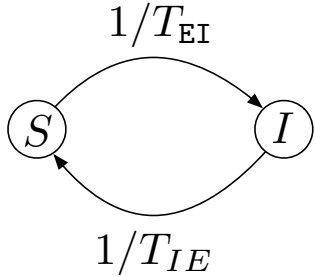}
 \caption{Flu state evolution of each agent.}
 \label{f:TPM-Flu}
\end{figure}

One final component of our model is the evolution of Flu state
$F_i(t)$. We remark that the inclusion of this state is only to model
the noise in applying a testing policy. Thus, we use a simple, intrinsic, SI model
for $F_i(t)$ where it is evolves as a Markov chain independently for each $i$
and taking values in $\{S, I\}$ with transition probabilities depicted in Figure~\ref{f:TPM-Flu}.
Note that an important distinction between the COVID and Flu models is that in the former the process ``terminates'' once the state $R$ is reached
since we assume that a person who has been infected with COVID-19 once cannot be infected again. However, a person keeps on shifting from $S$ to $I$
Flu states indefinitely -- this is because the "Flu state" is an proxy for the individual contracting any illness with similar symptoms as COVID-19. As we shall elaborate later,
any testing policy that makes a pool of symptomatic patients will treat an agent with $C_i(t)=I$ and $F_i(t)=I$ identically (if all their other features such as locality match). 

This template for an interaction model is very generic
and can incorporate various movements across the city in form of
origin-destination flow data. Even bus routes can be included by
considering them as locations. However, we have only implemented a
restricted form of this interaction in the first version of our
simulator used for experiments in this paper. Specifically, the number of fixed and
randomly selected agents met in the neighbourhood and the locality
visited are fixed to be the same for all the agents. Furthermore,
only one visited location is set per agent, as a part of its feature
matrix. However, we remark that it is easy to extend our model to
include multiple locations (such as workplace, bus-used, {\it
  etc.}). 

Before specifying the parameters used for the results of this paper in the next section,
we remark that our framework requires mobility data to be fed in form
of an {\em origin-destination} (OD) matrix. Such a matrix has number of
rows $N_r$ as the number of localities and the number of columns $N_c$
representing the locality or community visited daily.
The
$(i,j)$-th entry of this matrix represents the probability with which
a person from locality $i$ goes to locality $j$, whereby the $i$th row
constitutes a probability vector that sums to one. We use this vector
to generate the locality visited by each agent in locality $i$,
independently of other agents. Note that we can have multiple such OD
matrices, one for each community, and each person can be a member of
one community each for every matrix available in the model. In its current form, our
implementation allows the membership to depend only on the location
of the agents. Furthermore, agents have not been endowed with other potentially relevant  features
such as age and comorbidity which determine COVID state evolution in
practice. We plan to include these enhancements in a subsequent
version.

We close this section by noting that the state evolution of our model is defined in the file {\tt evolution.py}
of our implementation. Also, we point out that the state of the city is captured by a pandas data frame City Population, abbreviated as CP, with $N$ rows
and columns corresponding to various features of each agent. 

{\it A caution:} Our implementation is designed to use multiple CPU cores in parallel, but the specific
dynamics change when number of cores are increased. In particular,
increasing the number of cores used changes the number of state
transitions happening in parallel, reducing the overall growth
rate. Throughout this paper, we have set the number of cores to 8. 
\subsection{Parameters used for simulation}
\label{subsec:SimParams}
\begin{figure}[h]
  \centering
  \includegraphics[scale=0.7]{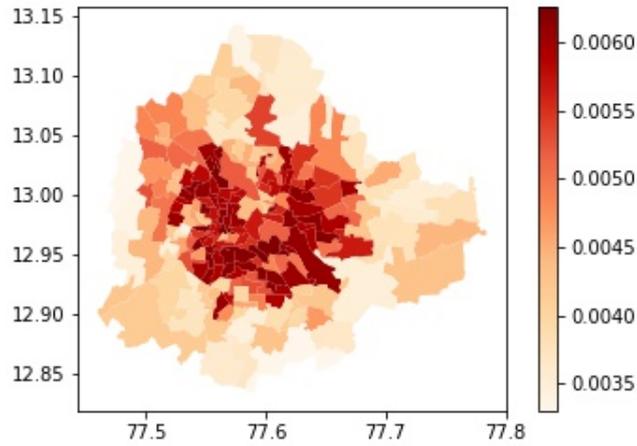}
 \caption{Ward-wise population density of Bengaluru.}
 \label{f:BangalorePopulation}
\end{figure}

In this paper, we instantiate the general framework described above
for the city of Bengaluru in India. For simplicity, we only consider
an SIR model for COVID state where the E state is skipped by setting
$T_{EI}=1$. We consider the 198 urban wards of Bengaluru that come
under the city municipal corporation (Bruhat  Bengaluru Mahanagara Palike, 
BBMP) and use the census data (converted to a {\tt .geojson} file using data
from~\cite{BangaloreGeoData,BangaloreDemographics}) to populate $N=100000$ agents across the
city; see Figure~\ref{f:BangalorePopulation} for depiction of
population densities. 

\begin{figure}[h]
  \centering \includegraphics[scale=0.3]{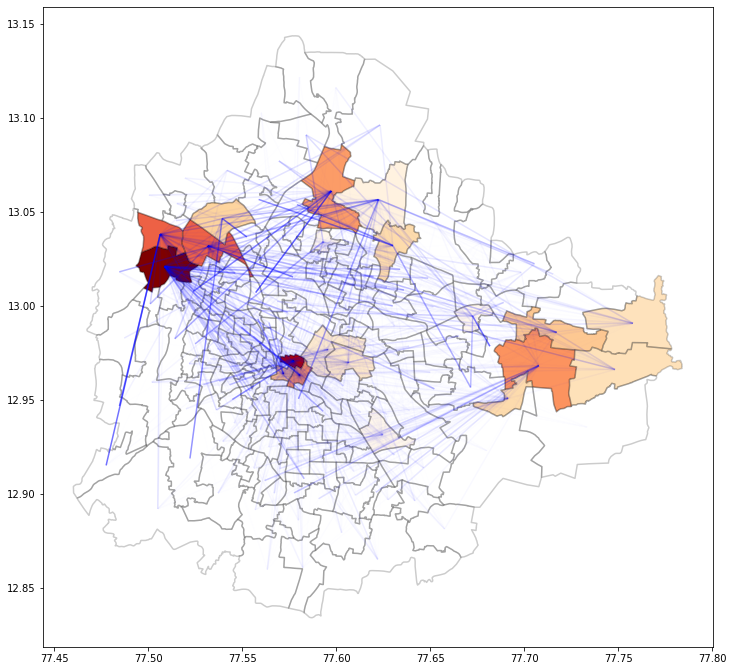}
  \caption{The top 20 wards with highest car traffic inflow in
    Bengaluru. The red component in the colour of each ward is proportional to the fraction of inflow traffic for the ward and the opaqueness of blue lines indicate the flow between the wards it joins.}
 \label{f:inflow}
\end{figure}

For modeling mobility across the city, we use data on vehicle mobility across Bengaluru
acquired from the Centre for Infrastructure, Sustainable Transportation and Urban Planning (CiSTUP),
Indian Institute of Science, Bangalore.
This is similar in spirit to the prescription in~\cite{Ferrettieabb6936}, but instead of dynamic digital data, we
use static data obtained using surveys similar to~\cite{Klepac2020.02.16.20023754,Lieabb3221}.
This data includes OD matrices for vehicles of different categories across Bengaluru which was obtained by conducting household surveys across the city. 
We have only used the data for car traffic in our interaction model.
Furthermore, it includes daily bus ticket sales obtained from BBMP; however, in this paper, we have not used the bus ticket sales data.
To reduce computational load, we restrict the OD matrix to the 20 destination
locations seeing the highest inflow, depicted in
Figure~\ref{f:inflow}.  Note that we also have a fictitious
destination 21 representing an agent not visiting any of these 20
destination.
With the interaction model set, we now list the values set for various
parameters of our model described in the previous section:
\begin{center}
  \begin{tabular}{|l|l|}
        \hline Parameter& Value \\ \hline COVID infection rate $p$ &
        0.1 \\ \hline Average time from E to I for COVID state
        $T_{EI}$ &1 \\ \hline Average time from I to R for COVID state
        $T_{IR}$ &8 \\ \hline Average time from S to I for Flu state
        $T_{SI}$ &50 \\ \hline Average time from I to S for Flu state
        $T_{IS}$ &8 \\ \hline Number of randomly selected people each
        person meets in its neighbourhood &1 \\ \hline Number of fixed
        people each person meets in its neighbourhood &5 \\ \hline
        Number of randomly selected people each person meets at its
        workplace locality &2 \\ \hline Number of fixed people each
        person meets at its workplace locality &10 \\ \hline
\end{tabular}
\end{center}
We remark that while our focus in this paper is not on careful calibration of the model,
we have used reasonable parameter values based on results in~\cite{he2020temporal,linton2020incubation}.
We execute each simulation for $1,00,000$ people for a duration of $T=100$
days.  We mention in passing that we can easily incorporate bus data
as well as movement data collected from location traces as suggested
by~\cite{Ferrettieabb6936} in our framework; we will do this in subsequent
versions of our implementation.

\subsection{The testing policy framework}
A strength of our framework is its ability to incorporate testing
policies. Our simulator takes as input a function {\tt testingPolicy}
described in the file {\tt tests.py}. The history of testing of entire
population is stored in an $(N\times T)$ matrix TestingHistory, which
stores $0$ as the default value for each entry and updates the $(i,t)$-th
entry to $1$ or $-1$, respectively, if agent $i$ is tested on day $t$
and the outcome is positive or negative.

We can admit any testing
policy that selects a pool of candidates to be tested and applies the
a common test function ${\tt test()}$ to each individual in the pool.The
test function is defined to have $0$ probability of a false positive (this is in line with observed characteristics of the RT-PCR test for COVID-19 \cite{rtpcr-err}),
but a nonzero probability of false negative can be set. For
simplicity, we have set the probability of false negative to $0$ in
our simulations\footnote{The noise in tests appear because of the
  presence of another flu with similar symptoms as COVID-19.}.

In selecting the pool of candidates, a testing policy
can only use observable features such as locality of individual, the
workplace visited, and whether an agent is symptomatic. Note that
a {\it symptomatic agent} can have either Flu state I or COVID state
I, but a testing policy (obviously so) cannot use the actual COVID or
Flu state of a person.

We have implemented and simulated three testing
policies. Each policy selects a fixed number $N_{\tt test}$ of
symptomatic agents at each time step (day) to apply tests. The following policies are simulated:
\begin{enumerate}
\item {\it Random Symptomatic Testing (RST):} The
  daily pool of agents to be tested is selected randomly (from among people who are symptomatic).

\item {\it Contact Tracing (CT):} The daily pool is selected randomly
  from the set of all symptomatic among the neighbourhood contacts
  and the workplace contacts of all the patients. If this set is of
  smaller cardinality than $N_{\tt test}$, which happens often, we
  enhance it with symptomatic from across locations.

\item {\it Location Based Testing (LBT):} This policy is a bit
  involved and will be described in detail in Section \ref{s:testing}. At a high level, this policy gives priority for testing to
  symptomatic agents belonging to localities and workplaces with
  higher infection.
\end{enumerate}
A few remarks are in order. First, these
policies are enabled by maintaining a list of fixed contacts in the
neighbourhood and the workplace, which can be identified for CT. We assume that the {\em randomly selected} daily contacts that an agent interacts
with cannot be traced, as is the situation in practice.

Second, we
mention that CT is an ``infra-heavy'' testing policy requiring
operational support to trace contacts and test them. In contrast, RST
and LBT are algorithms that do not require operational support but
require a careful design of policy mechanisms to ensure prescribed
sampling.

Finally, we briefly outline the connection between testing policies in
practice and the ones we have implemented in our simulation
framework. 
All the testing policies described above start with a list of
symptomatic agents. In practice, such a list emerges 
when symptomatic patients reach out to their medical provider. An
effective information campaign by the government can ensure that
patients with symptoms matching those seen in COVID-19 come out for
testing. Nonetheless, the exact percentage of symptomatic patients
that come out can vary with localities, since each locality differs in
income levels and availability of medical facilities. To model this,
we incorporate ``under-reporting'' in 
our framework which can be represented by a locality wise reporting
probability vector. Further, our tests sample randomly from the pool
of symptomatic patients. In practice, this sampling is implemented by
medical professionals who recommend a subset of symptomatic patients
for testing. We believe that all the policies we have simulated can
be (and, for some, have been)  converted to an implementable form on the 
ground at least in the Indian context, e.g., the containment plan of the Ministry of Health and Family Welfare, Govt. of India, already lays down ``hotspot detection' as part of a testing/intervention strategy \cite{mohfwHotspots}.

\subsection{The interventions framework}
Our framework can simulate evolution under interventions. A policy
intervention such as a city-wide lockdown is implemented as modifying the interactions
between different agents in the simulation. The intervention policy is specified as an
input in the form of Python function {\tt interventionPolicy}  which is
accessed everyday (for every iteration) to produce a list of
interventions. When updating the state of each agent (using the
function {\tt updateState} specified in {\tt evolution.py}), the list of
interventions enabled on each day is used to decide which interactions
will be allowed for the agent. This is done by making access to the
function {\tt InterventionRule},  which interprets the impact of
intervention for state evolution. In particular, we have implemented
and simulated the following three interventions available as
respectively as Python functions
{\tt InterventionQuarantine}, {\tt InterventionLockdown}, and {\tt
  InterventionLockdownFixed} specified in {\tt interventions.py}:

\begin{enumerate}
\item {\it Quarantine:} Each agent that is tested positive for
  COVID-19 on day $t$, along with all the
  agents on its list of neighbourhood and workplace contacts, are placed
  on quarantine for a period of 10 days starting from day
  $t+1$. Namely, the agent is not allowed to interact with any other
  agent during this quarantine period.

\item {\it Indefinite lockdown:}  All agents are not allowed to
  interact with any other agent once a particular trend is detected
  in the positive test count.

\item {\it Fixed duration lockdown:}  All agents are not allowed to
  interact with any other agent for a fixed period of time once a
  particular trend is detected in the positive test count.
\end{enumerate}  
In the last two lockdowns, the rule to start a lockdown has not explicitly been mentioned. We will elaborate on this later, but roughly, a lockdown is triggered once the slope of positive COVID-19
tests crosses a threshold. Thus, the efficacy of an intervention policy is
tied intrinsically to the testing policy used. As such the intervention
policy function is given access to the overall population state CP,
the history of interventions imposed, and the testing history
matrix. If an additional state is required for implementing a policy
-- such as quarantine requires us to store the quarantine state of
each agent -- it is included as a part of CP.

In summary, our simulation framework separates state evolution
functions (in {\tt evolution.py}), testing policy (in {\tt tests.py}),
and intervention policy (in {\tt interventions.py}). A new test or a
new intervention can easily be incorporated by maintaining the
input-output structure.

\section{Testing Strategies and Performance Comparison}
\label{s:testing}
This section details the testing strategies that we have explored
using the simulation framework of Section \ref{s:simulation}, and
compares their performance. For this section,
no intervention is performed based on the test outcomes. The impact of
tying interventions to test outcomes is discussed in Section
\ref{s:interventions}.     

\subsection{Testing strategies} At a high level, a test strategy is
expressed as a {\em test selection rule}, applied in each time step
(day) of the simulation, that maps the current population, together
with the past testing history, to a subset of individuals that are
subsequently tested for COVID-19 infection. This map cannot depend on
the actual health states of the individuals (e.g., their COVID state),
but can rely on only their {\em observable} attributes, such as
whether they display symptoms of illness or whether they reside in a
specific ward or wards. The map can also be randomized to reflect
random sampling from certain locations without necessarily relying on
the onset of symptoms. The general pseudocode of a test selection rule
appears in Algorithm~\ref{alg:testselection}.   

All individuals selected by a test selection rule are assumed to
undergo medical testing (e.g., RT-PCR testing) represented by the {\em
  individual test} subroutine Algorithm \ref{alg:testindiv} (defined as the function {\tt test} in {\tt tests.py}), which
models false negatives arising in the testing process at an assumed
rate $r \in [0,1]$.  

In the sequel, the testing strategies we describe are specifications
of test selection rules, assuming access to a standard individual
testing subroutine.

\begin{algorithm}[H]
\SetAlgoLined
\SetKwInOut{Input}{input}\SetKwInOut{Output}{output}
\Input{Current day $t$, Past testing history $\mathcal H_{t-1}$ ,
  i.e., individuals tested on day $s \leq t-1$ and their test results,
  Observable attributes of all individuals on day $t$} 
\Output{Set of individuals to be tested on day $t$}  
\BlankLine
$S = \emptyset$ \tcp*[r]{holds set of individuals to be tested}
\For{individual in population}{
\If{\emph{condition} is satisfied}{
add individual to $S$ \;}
}

\KwRet{$S$}

 \caption{Test selection algorithm (general)}
 \label{alg:testselection}
\end{algorithm}

\begin{algorithm}[H]
\SetAlgoLined
\SetKwInOut{Input}{input}\SetKwInOut{Output}{output}
\Input{Individual $i$, False negative rate $r$}
\Output{Result $\in \{+1, -1\}$}  
\BlankLine
Result $\leftarrow -1$\;
\If{$i$'s CovidState is ``I"}{
	Result $\leftarrow +1$ with probability $(1-r)$
}
\KwRet{Result}

\caption{TestIndividual (Individual test subroutine)}
\label{alg:testindiv}
\end{algorithm}

\subsubsection{Random Symptomatic Testing}

The first, and simplest, testing strategy we consider is {\em Random Symptomatic Testing (RST)}. This strategy (pseudocode in Algorithm \ref{alg:RST}) looks at the pool of people who display symptoms typical of the disease, and randomly samples as many of them as possible to fill up a predefined budget. This is considered a baseline testing strategy under budget constraints in the experiments reported here.
 
\begin{algorithm}[H]
\SetAlgoLined
\SetKwInOut{Input}{input}\SetKwInOut{Output}{output}
\Input{{\bf Test budget $b$}, Current day $t$, Past testing history $\mathcal H_{t-1}$ , i.e., individuals tested on day $s \leq t-1$ and their test results, Observable attributes of all individuals on day $t$}
\Output{Set of individuals to be tested on day $t$}  
\BlankLine

$A \leftarrow $ Individuals displaying symptoms on day $t$ \;

$S \leftarrow $ Random subset of $\min(b, |A|)$ individuals, sampled from $A$ without replacement \;

\KwRet{$S$}

 \caption{Random Symptomatic Testing (RST) test selection algorithm}
 \label{alg:RST}
\end{algorithm}

\subsubsection{Contact Tracing}
Contact Tracing (CT) is a testing strategy in which symptomatic contacts of
individuals that have tested positive in the recent past\footnote{In
  our implementation of CT, the recent past is taken to be the past 2
  days.} are tested at the highest priority. If any more tests are available, then as many (randomly
chosen) individuals as possible that are presently exhibiting symptoms
are chosen for testing. Pseudocode for the CT test selection strategy
appears in Algorithm \ref{alg:CT}. Note that in practice even the
nonsymptomatic contacts are often tested, which is what should be done
in a simulation with the SEIR model. But since we have disabled the
`E' (exposed)
state for simplicity, we only test symptomatic contacts.

\begin{algorithm}
\SetAlgoLined
\SetKwInOut{Input}{input}\SetKwInOut{Output}{output}
\Input{{\bf Test budget $b$}, Current day $t$, Past testing history $\mathcal H_{t-1}$ , i.e., individuals tested on day $s \leq t-1$ and their test results, Observable attributes of all individuals on day $t$}
\Output{Set of individuals to be tested on day $t$}  
\BlankLine

$A \leftarrow $ Individuals displaying symptoms on day $t$ \;

$C \leftarrow$ All (fixed) symptomatic contacts of individuals tested +ve in the past 2 days (day-1 \& day-2)

\uIf(\tcp*[l]{more tests available than \# traced contacts}){$|C| < b$}{
$S \leftarrow C \cup \left\{\min\{b-|C|, |A|\} \text{ individuals sampled from $A$ without replacement} \right\} $  
}
\Else(\tcp*[l]{fewer tests than \# traced contacts}){
$S \leftarrow \left\{ b \text{ individuals sampled from $C$ without replacement} \right\}$
}

\KwRet{$S$}

 \caption{Contact Tracing (CT) test selection algorithm}
 \label{alg:CT}
\end{algorithm}

\subsubsection{Location-Based Testing}
Location-Based Testing (LBT) is a testing rule that is designed to favour individuals who are `close' to the currently known footprint of the COVID-19 infection. Closeness here is assumed to be high if (a) the individual's locality contains many individuals known to have tested +ve in the past, or (b) many individuals who have tested +ve in the past are associated with the individual's visit place.

The LBT selection rule (Algorithm \ref{alg:LBT}) essentially computes a closeness or risk score of each person who reports symptoms, and prioritises individuals for testing depending on their risk scores. The risk score of a person can be thought of as a crude proxy for the posterior probability of that person being infected with COVID-19 on a given day, given all the observed history of tests. 

For our implementation, we define the score of an individual $i$ on day $t$ as a weighted sum of the scores of its residence locality and its visit place. The score of a locality is an exponentially weighted average of the number of +ve tested individuals associated to it in the past, e.g., an individual who tested positive from the locality $\Delta$ days ago contributes $\alpha_{\text{loc}} (1+\epsilon)^{\Delta-1}$ to that locality's score where $\alpha_{\text{loc}} > 0$ is an adjustable parameter. The score of a visiting place is defined analogously. Pseudocode for the LBT selection algorithm appears in Algorithms \ref{alg:LBT} and \ref{alg:getScore}.

\begin{algorithm}
\SetAlgoLined
\SetKwInOut{Input}{input}\SetKwInOut{Output}{output}
\Input{{\bf Test budget $b$}, Current day $t$, Past testing history $\mathcal H_{t-1}$ , i.e., individuals tested on day $s \leq t-1$ and their test results, Observable attributes of all individuals on day $t$}
\Output{Set of individuals to be tested on day $t$}  
\BlankLine

$A \leftarrow $ Individuals displaying symptoms on day $t$ \;

\For{Person in $A$}{
Score[$i$] $\leftarrow$ getScore($i$, $t$)
}

$S \leftarrow$ Random subset of $\min(b, |A|)$ individuals from $|A|$, successively sampled without replacement according to their Score

\KwRet{$S$}

 \caption{Location-Based Testing (LBT) test selection algorithm}
 \label{alg:LBT}
\end{algorithm}

\begin{algorithm}
\SetAlgoLined
\SetKwInOut{Input}{input}
\SetKwInOut{Output}{output}
\SetKwInOut{Params}{parameters}
\Input{Individual $i$, Day $t$}
\Output{(Risk) Score of $i$} 
\Params{$\alpha_{\text{loc}}$ (weight for locality per +ve case), $\alpha_{\text{vis}}$ (weight for visit place per +ve case), $\beta$ (relative weight of locality w.r.t. visit place), $\epsilon$ (amplification factor)} 
\BlankLine

$v \leftarrow$ Visiting place which  $i$ visits \;

$\ell \leftarrow$ Locality where $i$ resides \;

Score\_$v$ $\leftarrow 0$ \;

Score\_$\ell$ $\leftarrow 0$ \;

\For{$\tau$ in $1:(t-1)$}{
Score\_$v$ $\leftarrow$ $(1 + \epsilon)$Score\_$v$ +  $\alpha_{\text{loc}}  \left|\left\{ j: j \text{ tested +ve on day $\tau$ \& $j$ visits $v$}  \right\} \right|$ \;
Score\_$\ell$ $\leftarrow$ $(1 + \epsilon)$Score\_$\ell$ +  $\alpha_{\text{vis}}  \left|\left\{ j: j \text{ tested +ve on day $\tau$ \& $j$ resides in $\ell$}  \right\} \right|$

}

\KwRet{Score\_$v$ + $\beta$ Score\_$\ell$}

 \caption{getScore subroutine}
 \label{alg:getScore}
\end{algorithm}

\noindent {\bf Remark.} We have not modelled the fact that both the
conduct of tests and their reporting can suffer delays. For the
interested experimenter, this can  be incorporated easily by having
the result of the  TestIndividual subroutine (Algorithm
\ref{alg:testindiv}) available to the parent test selection routine
after a suitable delay.

\subsection{Numerical Results and Discussion}

We present and contrast here the numerical performance of the 3 testing strategies described previously -- Random Symptomatic Testing (RST), Contact Tracing (CT) and Location-Based Testing (LBT). We run our experiments using the parameter settings of  Section \ref{subsec:SimParams}, without any nontrivial interventions enabled for individuals who test positive for COVID-19. 

\subsubsection{Test Performance with Clustered Seeding} 
Clustered seeding simulates the initial condition where all the initial COVID-19 cases are spatially localized. In our experiments, we initialize 50 COVID-19-infected individuals (out of a population of 100k) in a single locality (ward
number 120, the `Cottonpete ward') of the city of Bengaluru and let the outbreak evolve from there.  

\begin{figure}[ht]
 \begin{subfigure}{.5\textwidth}
   \centering
  \includegraphics[width=.8\linewidth]{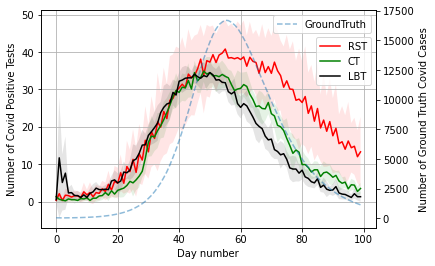}
    \caption{Daily number of tests.}
\end{subfigure}
 \begin{subfigure}{.5\textwidth}
      \centering
  \includegraphics[width=.8\linewidth]{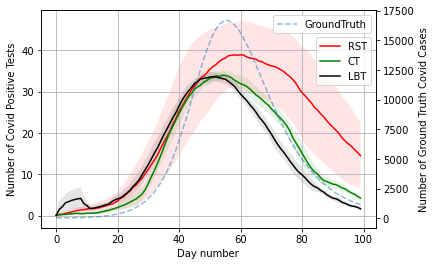}
  \caption{Daily number of tests smoothened by taking an average over a window of 8 days.}
\end{subfigure}
\caption{Comparative test performance with {\bf clustered seeding} and without intervention, for a time period of 100 days and with a testing budget of 50 tests/day. Results are averaged across 10 runs and error bars represent 1 standard deviation.}
\label{f:NUseed_IntervNone}
\end{figure}

Figure \ref{f:NUseed_IntervNone} depicts the mean daily test score
(number of tests with positive results) evolution with time for 10
independent simulation runs, along with the corresponding 1-standard
deviation ranges shaded in lighter colour, for a daily budget of 50
tests/day. The ground truth number of cases per day is also plotted in
the background. It is interesting to note that (a) RST shows high
variance in reporting compared to the more biased CT and LBT, (b)
while all 3 tests are equally accurate in capturing the trend of the
ground truth in the phase leading up to the peak of actual case count,
their performance after the peak is reached is quite different -- CT
and LBT tend to fall earlier than RST\footnote{We 
  have not formally defined the notion of ``trend,'' but a good proxy
  to keep in mind is the derivative of the ground truth curve, which
  seems to be well approximated by the derivative of the positive
  tests curve. This notion of approximation is well studied in
  nonparametric statistics, and can be used for a theoretical
  treatment of COVID-19 ground truth estimation.}.  Also interesting
is the short-lived peak in the curve for LBT early on -- this is due
to the test aggressively prioritising testing from the affected seed
locality. A smoothed version of the test results, plotted alongside,
shows that smoothing can make the outputs of CT and LBT capture the
rise-fall trend of the ground truth in a better fashion. Note that we
have chosen a smoothing window of 8 days since the average time from
COVID state 'I' to COVID state 'R' in our model is 8 days.
Presumably the gains due to smoothing can be attributed to this fact
-- a patient tested positive 8 days ago is likely to remain positive
till the current day.

\begin{figure}[ht]
 \begin{subfigure}{.5\textwidth}
   \centering
  \includegraphics[width=.8\linewidth]{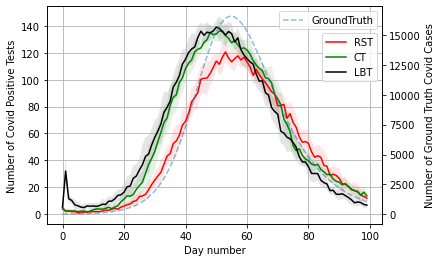}
    \caption{Daily number of tests.}
\end{subfigure}
 \begin{subfigure}{.5\textwidth}
      \centering
  \includegraphics[width=.8\linewidth]{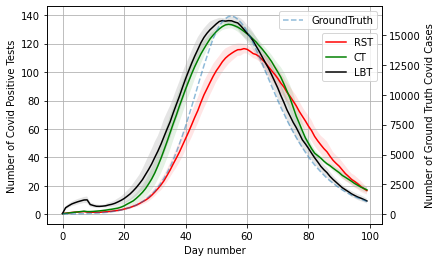}
  \caption{Daily number of tests smoothened by taking an average over a window of 8 days.}
\end{subfigure}
\caption{Comparative test performance with clustered seeding and without intervention, for a time period of 100 days and with a {\bf testing budget of 200 tests/day}. Results are averaged across 10 runs and error bars represent 1 standard deviation.}
\label{f:NUseed200_IntervNone}
\end{figure}

We also compare the tests with an enhanced (4x) budget of 200
tests/day (Figure \ref{f:NUseed200_IntervNone}). The extra budget
appears to put to good use by the `smart' tests CT and LBT, whose test
numbers outstrip those of RST by a significant margin in the lead up
to the peak. The faithfulness to the actual ground truth signal also
appears to be much better for CT and RST here.

\subsubsection{Test Performance with Uniform Seeding} 

\begin{figure}[ht]
 \begin{subfigure}{.5\textwidth}
   \centering
  \includegraphics[width=.8\linewidth]{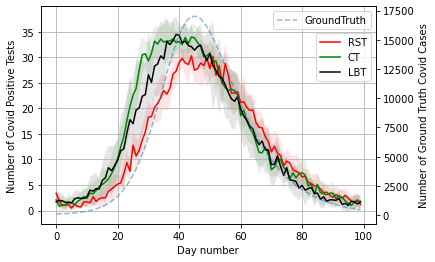}
    \caption{Daily number of tests.}
\end{subfigure}
 \begin{subfigure}{.5\textwidth}
      \centering
  \includegraphics[width=.8\linewidth]{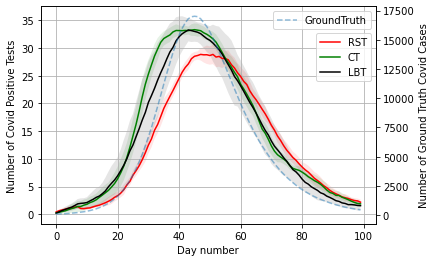}
  \caption{Daily number of tests smoothened by taking an average over a window of 8 days.}
\end{subfigure}
\caption{Comparative test performance with {\bf uniform seeding} and without intervention, for a time period of 100 days and with a testing budget of 50 tests/day. Results are averaged across 10 runs and error bars represent 1 standard deviation.}
\label{f:Useed_IntervNone}
\end{figure}

\begin{figure}[ht]
 \begin{subfigure}{.5\textwidth}
   \centering
  \includegraphics[width=.8\linewidth]{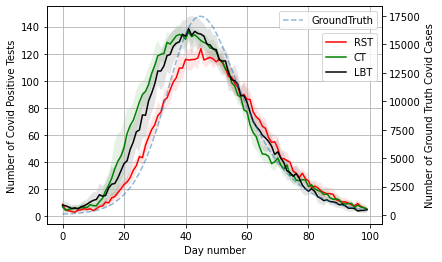}
    \caption{Daily number of tests.}
\end{subfigure}
 \begin{subfigure}{.5\textwidth}
      \centering
  \includegraphics[width=.8\linewidth]{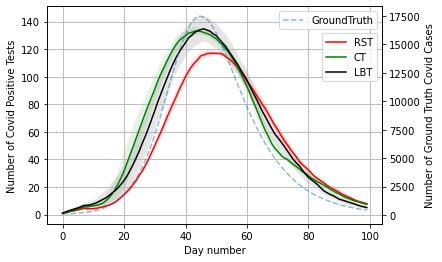}
  \caption{Daily number of tests smoothened by taking an average over a window of 8 days.}
\end{subfigure}
\caption{Comparative test performance with uniform seeding and without intervention, for a time period of 100 days and with a {\bf testing budget of 200 tests/day}. Results are averaged across 10 runs and error bars represent 1 standard deviation.}
\label{f:Useed200_IntervNone}
\end{figure}

Figure \ref{f:Useed_IntervNone} shows the results of applying the 3 tests with an initial seeding that is {\em uniform} across localities (city wards). Specifically, each of the (approx.) 200 localities in the simulation model hosts an independent  Binomial$(5, 0.1)$ number of COVID-19 infected seeds at start, resulting in about 100 seeds in the overall population (100k). Figure \ref{f:Useed200_IntervNone} plots the same metrics but for tests with an enhanced testing budget (200 per day).

The results show a clear advantage of the more advanced tests (CT, LBT) over the RST baseline, in the period leading up to the peak of ground truth COVID-19 cases. This potentially brings out the value of relying on predictive biased sampling (over and above the symptomatic sampling by RST) to detect a higher number of cases in the initial stages of the outbreak. We will see later (Section \ref{s:interventions} ) that this confers a significant advantage in terms of timing when the results of the former tests are used to implement public (large-scale) lockdowns.

\subsubsection{Test Performance with Under-reporting of Symptoms}

We also examine what happens when individuals under-report symptoms when ill. This has the effect, in our simulation, of reducing the initial pool of symptomatic individuals which are input to the test selection strategy. 

\begin{figure}[ht]
 \begin{subfigure}{.5\textwidth}
   \centering
  \includegraphics[width=.8\linewidth]{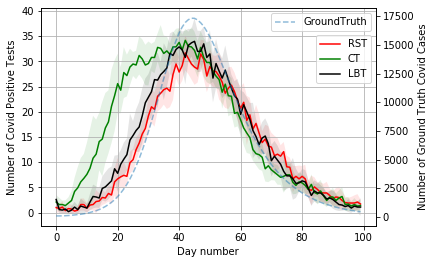}
    \caption{Daily number of tests.}
\end{subfigure}
 \begin{subfigure}{.5\textwidth}
      \centering
  \includegraphics[width=.8\linewidth]{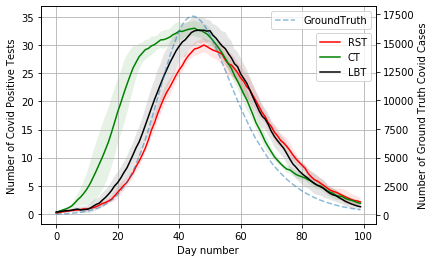}
  \caption{Daily number of tests smoothened by taking an average over a window of 8 days.}
\end{subfigure}
\caption{Comparative test performance with uniform seeding and without intervention, for a time period of 100 days, a testing budget of 50 tests/day, and {\bf 10\% reporting of symptoms}. Results are averaged across 10 runs and error bars represent 1 standard deviation.}
\label{f:UseedUnderreport_IntervNone}
\end{figure}

{\bf Uniform under-reporting.}  Figure \ref{f:UseedUnderreport_IntervNone} shows the test performance curves when each individual who is infected (either with COVID-19 or "flu") is assumed to report as symptomatic with probability $0.1$ independently (in the previous sections, this was assumed to occur with probability $1$). Though the 10x underreporting does not significantly affect the way in which all 3 tests capture the rise and fall in ground-truth cases, contact tracing emerges as the most informative detector of cases in the lead up to the peak. 

\begin{figure}[ht]
 \begin{subfigure}{.5\textwidth}
   \centering
  \includegraphics[width=.8\linewidth]{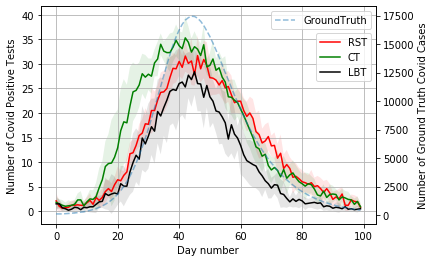}
    \caption{Daily number of tests.}
\end{subfigure}
 \begin{subfigure}{.5\textwidth}
      \centering
  \includegraphics[width=.8\linewidth]{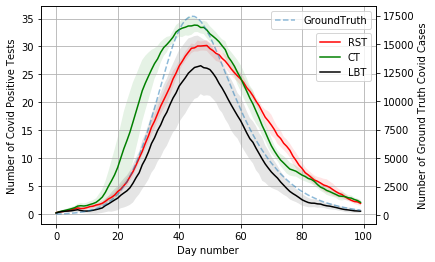}
  \caption{Daily number of tests smoothened by taking an average over a window of 8 days.}
\end{subfigure}
\caption{Comparative test performance with uniform seeding and without intervention, for a time period of 100 days, a testing budget of 50 tests/day, and {\bf non-uniform reporting of symptoms}. Results are averaged across 10 runs and error bars represent 1 standard deviation.}
\label{f:UseedNUUnderreport_IntervNone}
\end{figure}

{\bf Non-uniform under-reporting.} Figure \ref{f:UseedNUUnderreport_IntervNone} depicts the tests operating in a scenario where individuals report in a non-uniform manner whether they are symptomatic and hence to be considered for testing. Specifically, individuals from about $1/3$rd of the localities (wards) in the city (selected at random) report symptoms at rate 5\% while the rest report at rate 100\%. The initial seeding for COVID-19 cases is uniform across the localities as before. Contact tracing is seen to be robust to the rather skewed under-reporting enforced here, presumably due to its more accurate predictions of infection targets due to a richer local signal.


\subsubsection{Estimating the Ground Truth number of Infections}
\begin{figure}[ht]
   \centering
  \includegraphics[width=.5\linewidth]{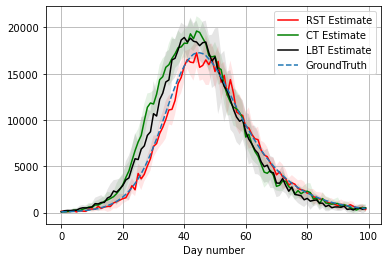}
\caption{Estimated no. of COVID-19 cases computed using numbers of symptomatic (at 100\% reporting) and positive case counts. The simulation was carried out with  uniform seeding and no intervention and a testing budget of 50 tests/day.}
\label{f:UseedEstimate}
\end{figure}

We explore the question of how well the test results, along with a
record of the number of symptomatic individuals considered for
testing, can be used to estimate the ground truth number of COVID-19
infections at any point in time. To this end, Figure
\ref{f:UseedEstimate} plots, for each day, an estimate of the ground
truth computed as: the number of symptomatic patients $\times$ the
error rate of the test, where the error rate of the test (on a day) is
the ratio of the number of positive tests to the number of total tests
performed. It is observed from the plot that the the (almost) unbiased
RST algorithm gives the best fit to the ground truth evolution. The
more aggressive (biased) CT and LBT strategies tend to overestimate
the ground truth during the initial stages of the epidemic. We do not
consider the problem of forming a more precise estimate using these
strategies -- it will require a precise knowledge of sampling
probabilities and it is unclear if that will be available in practice.

\subsection{Visualization through Geo Plots}
\begin{figure}[ht]
\centering
 \begin{subfigure}{.18\textwidth}
   \centering
  \includegraphics[width=\textwidth]{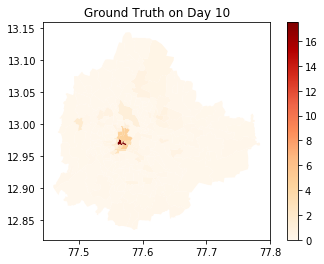}
\end{subfigure}
\hfill
 \begin{subfigure}{.18\textwidth}
   \centering
  \includegraphics[width=\textwidth]{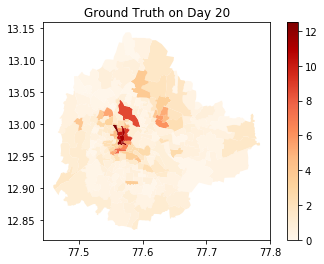}
\end{subfigure}
\hfill
 \begin{subfigure}{.18\textwidth}
   \centering
  \includegraphics[width=\textwidth]{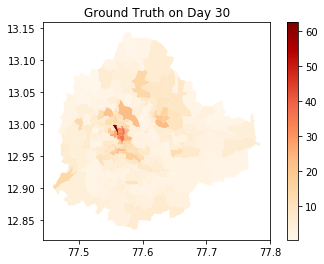}
\end{subfigure}
\hfill
 \begin{subfigure}{.18\textwidth}
   \centering
  \includegraphics[width=\textwidth]{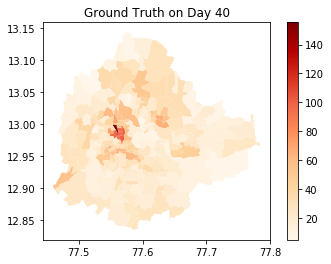}
\end{subfigure}
\hfill
 \begin{subfigure}{.18\textwidth}
   \centering
  \includegraphics[width=\textwidth]{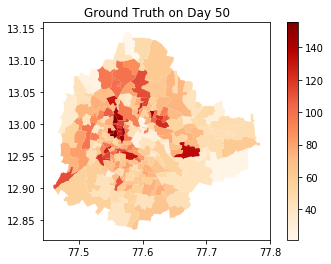}
 \end{subfigure}
 
 \begin{subfigure}{.18\textwidth}
   \centering
  \includegraphics[width=\textwidth]{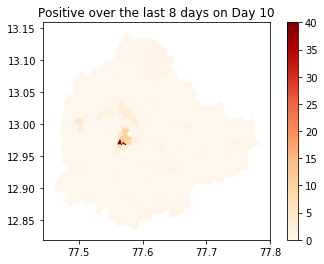}
\end{subfigure}
\hfill
 \begin{subfigure}{.18\textwidth}
   \centering
  \includegraphics[width=\textwidth]{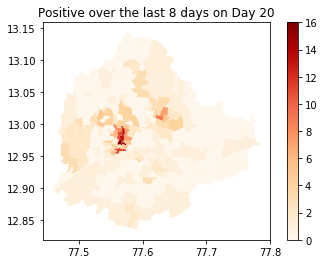}
\end{subfigure}
\hfill
 \begin{subfigure}{.18\textwidth}
   \centering
  \includegraphics[width=\textwidth]{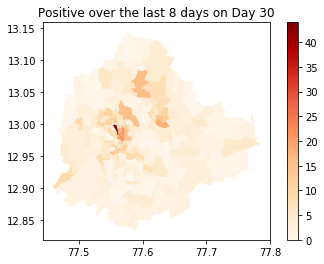}
\end{subfigure}
\hfill
 \begin{subfigure}{.18\textwidth}
   \centering
  \includegraphics[width=\textwidth]{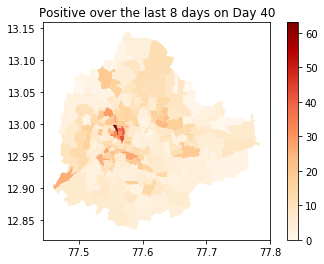}
\end{subfigure}
\hfill
 \begin{subfigure}{.18\textwidth}
   \centering
  \includegraphics[width=\textwidth]{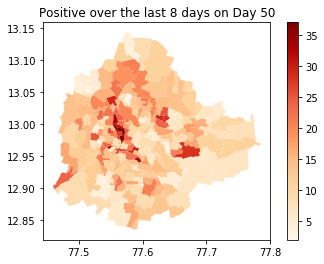}
 \end{subfigure}
 
 \begin{subfigure}{.18\textwidth}
   \centering
  \includegraphics[width=\textwidth]{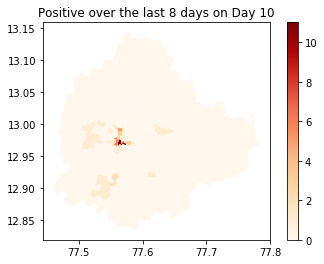}
\end{subfigure}
\hfill
 \begin{subfigure}{.18\textwidth}
   \centering
  \includegraphics[width=\textwidth]{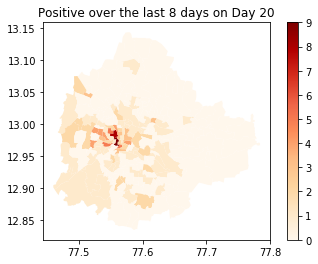}
\end{subfigure}
\hfill
 \begin{subfigure}{.18\textwidth}
   \centering
  \includegraphics[width=\textwidth]{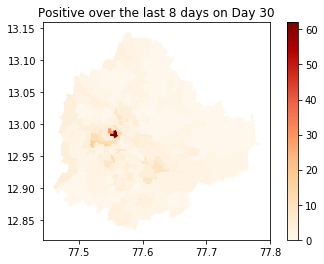}
\end{subfigure}
\hfill
 \begin{subfigure}{.18\textwidth}
   \centering
  \includegraphics[width=\textwidth]{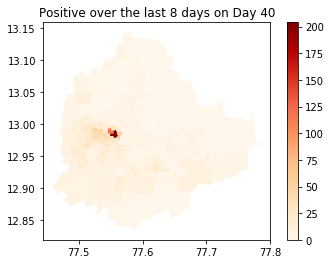}
\end{subfigure}
\hfill
 \begin{subfigure}{.18\textwidth}
   \centering
  \includegraphics[width=\textwidth]{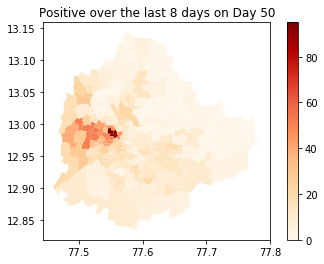}
 \end{subfigure}

 \begin{subfigure}{.18\textwidth}
   \centering
  \includegraphics[width=\textwidth]{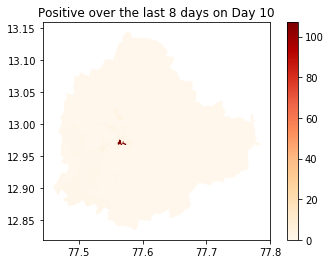}
\end{subfigure}
\hfill
 \begin{subfigure}{.18\textwidth}
   \centering
  \includegraphics[width=\textwidth]{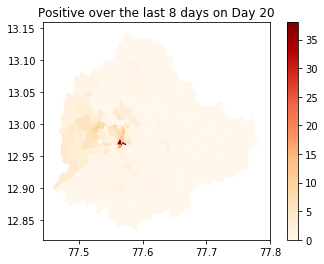}
\end{subfigure}
\hfill
 \begin{subfigure}{.18\textwidth}
   \centering
  \includegraphics[width=\textwidth]{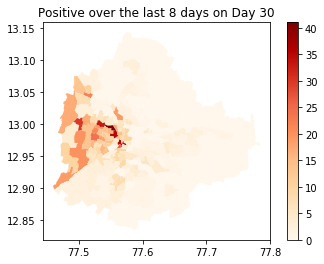}
\end{subfigure}
\hfill
 \begin{subfigure}{.18\textwidth}
   \centering
  \includegraphics[width=\textwidth]{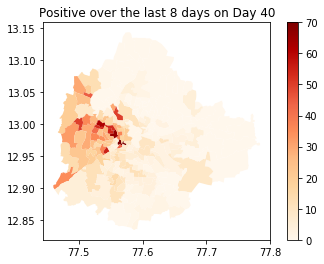}
\end{subfigure}
\hfill
 \begin{subfigure}{.18\textwidth}
   \centering
  \includegraphics[width=\textwidth]{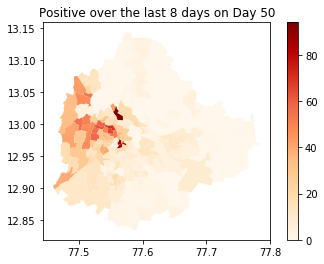}
 \end{subfigure}

\caption{Geo plots illustrating the evolution of the ground truth and positive cases across the city and over time, with uniform seeding of COVID-19 cases. The X-axis represents time, with each successive column denoting an increment of 10 days from the previous column (or day $0$ for the first column). The first row represents the heat map for simulated ground truth COVID-19 cases by locality in Bengaluru. The second, third and fourth rows represent the heat maps for COVID-19 positive cases detected per locality by the Randomized Symptomatic Testing (RST), Contact Tracing (CT) and Location-based Testing (LBT) test selection algorithms, respectively, for the past 8 days.}
\label{f:UseedGeo}
\end{figure}

We depict an example spatio-temporal evolution of ground truth
COVID-19 cases across the city, along with the corresponding  positive
cases detected in the recent past, in Figure \ref{f:UseedGeo}. This
illustrates the spatio-temporal nature of our simulation model and can
be useful for making qualitative inferences about certain testing
and/or intervention strategies.  It can be noted that while all tests
return a good signature of the source of infection initially,
LBT yields vey high number of positive tests during the initial
period, making it a suitable testing strategy for early detection and
containment of the spread.

\section{Effect on Interventions}
\label{s:interventions}
In this section, we compare the efficacy of our three testing policies
in enabling interventions. As outlined earlier, an intervention policy
takes into account the number of positive tests produced by the
testing policy. An intervention such as lockdown is enabled once a
particular trend is detected. Note that the purpose of such a lockdown is to ``flatten the curve" by reducing the maximum number of daily COVID-19 cases in the ground truth. Such a reduction will reduce the stress on hospitals and other medical facilities.

We simulated COVID-19 evolution using the
same rule for engaging an intervention policy using our testing
polcies of RST, CT, and LBT. All the results we present in this section represent average over 10 runs of simulation;
the mean behavior is depicted prominantly and the spread up to standard deviation is shown in a lighter color.  
Briefly, we find that CT outperforms RST
significantly in reducing the maximum number daily of COVID-19 cases (in the
ground truth). Interestingly, LBT performs comparably with CT --
recall that the former is less operations-intensive that the latter. 
We present the three interventions we have considered in separate
sections below.

\subsection{Quarantine}
\begin{figure}[h]
 \begin{subfigure}[h]{.5\textwidth}
   \centering
  \includegraphics[width=.85\linewidth]{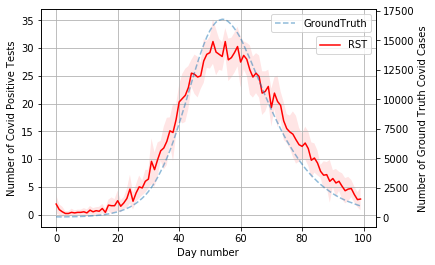}
    \caption{Daily number of tests.}
\end{subfigure}
 \begin{subfigure}[h]{.5\textwidth}
      \centering
  \includegraphics[width=.85\linewidth]{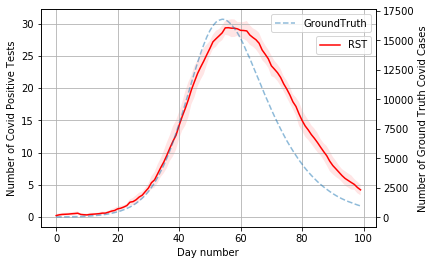}
  \caption{Daily number of tests smoothened by taking an average over a window of 8 days.}
\end{subfigure}
\caption{Evolution of COVID-19 cases under quarantine intervention
  using RST.}
\label{f:quarantine_RST}
\end{figure}

\begin{figure}[h]
 \begin{subfigure}[h]{.5\textwidth}
   \centering
  \includegraphics[width=.85\linewidth]{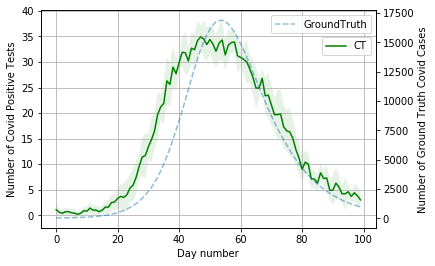}
    \caption{Daily number of tests.}
\end{subfigure}
 \begin{subfigure}[h]{.5\textwidth}
      \centering
  \includegraphics[width=.85\linewidth]{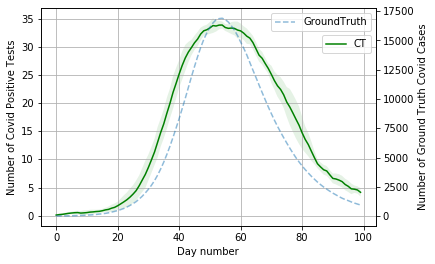}
  \caption{Daily number of tests smoothened by taking an average over a window of 8 days.}
\end{subfigure}
\caption{Evolution of COVID-19 cases under quarantine intervention
  using CT.}
\label{f:quarantine_CT}
\end{figure}

\begin{figure}[h]
 \begin{subfigure}[h]{.5\textwidth}
   \centering
  \includegraphics[width=.85\linewidth]{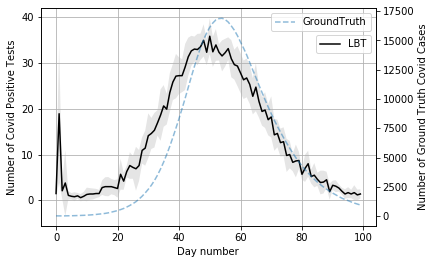}
    \caption{Daily number of tests.}
\end{subfigure}
 \begin{subfigure}[h]{.5\textwidth}
      \centering
  \includegraphics[width=.85\linewidth]{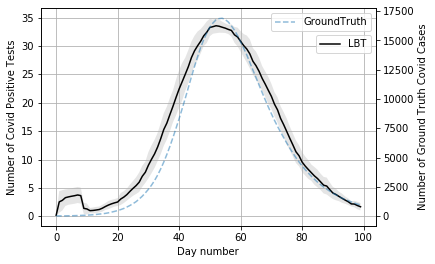}
  \caption{Daily number of tests smoothened by taking an average over a window of 8 days.}
\end{subfigure}
\caption{Evolution of COVID-19 cases under quarantine intervention
  using LBT.}
\label{f:quarantine_LBT}
\end{figure}

We consider quarantine intervention where any person tested
positive and all its contacts are placed under quarantine for 10
days. Note that this period of 10 days is shorted than the usual 14 to 21 days quarantine imposed in India; the reduction is to compensate for the
absence of the E state in our simplified SIR model (average E to I period observed is 5 days).
We have considered 50 tests  (1 test per 2000 agents) per day starting with a clustered
seeding where all the initial infections of COVID-19 are placed in one
locality. Specifically, we place 50 agents in COVID state I in ward
number 120 (the `Cottonpete ward'), as in the previous section. We
present our results using RST, CT, and LBT in
Figures~\ref{f:quarantine_RST},~\ref{f:quarantine_CT},
and~\ref{f:quarantine_LBT}, respectively.

We observe that since the number of tests is very few, quarantine
alone does not significantly reduce the maximum daily number COVID-19
cases (in the ground truth). Also, the number of positive tests that result using CT exceeds both RST and LBT,
while RST captures the derivative of the ground truth COVID-19 cases well. These observations are similar to those for the case when no intervention was done and can be attribued to the unbiasedness of RST estimates when all symptomatic patients come out for testing.

\begin{remark}
  Note that LBT gets high number of positive tests in the beginning. This phenomenon was seen for clustered seed in the previous section as well and can be attributed to the fact that LBT tests many symptomatic agents from the location where infection was initially seeded. Later, as the infection spreads across the city and LBT has captured a large fraction of the initial 50 cases in ward number 120, LBT starts performing comparably with CT. In fact, this observation will hold in all the other interventions as well. This feature of LBT can be exploited to enforce a local lockdown of ward number 120 early on, preventing the further spread of COVID-19. However, we do not consider this intervention in this paper and will visit it in follow-up work. 
  \end{remark}
\subsection{Indefinite lockdown}
\begin{figure}[h]
 \begin{subfigure}[h]{.5\textwidth}
   \centering
  \includegraphics[width=.85\linewidth]{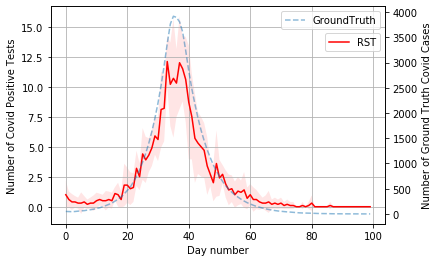}
    \caption{Daily number of tests.}
\end{subfigure}
 \begin{subfigure}[h]{.5\textwidth}
      \centering
  \includegraphics[width=.85\linewidth]{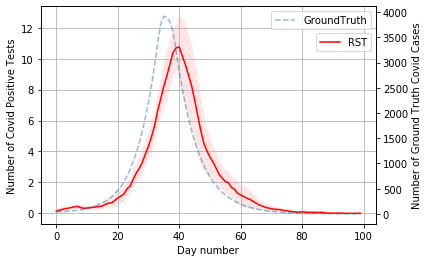}
  \caption{Daily number of tests smoothened by taking an average over a window of 8 days.}
\end{subfigure}
\caption{Evolution of COVID-19 cases under the indefinite lockdown intervention using the thresholded smoothened slope policy.}
\label{f:lockdown_RST}
\end{figure}

\begin{figure}[h]
 \begin{subfigure}[h]{.5\textwidth}
   \centering
  \includegraphics[width=.85\linewidth]{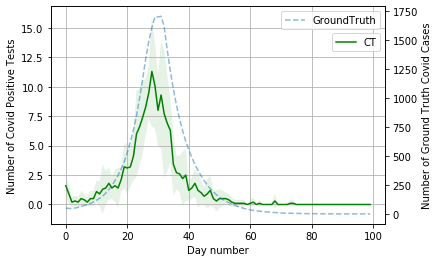}
    \caption{Daily number of tests.}
\end{subfigure}
 \begin{subfigure}[h]{.5\textwidth}
      \centering
  \includegraphics[width=.85\linewidth]{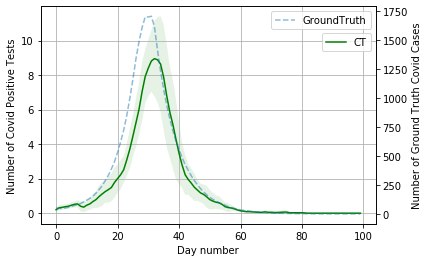}
  \caption{Daily number of tests smoothened by taking an average over a window of 8 days.}
\end{subfigure}
\caption{Evolution of COVID-19 cases under the indefinite lockdown intervention using the thresholded smoothened slope policy.}
\label{f:lockdown_CT}
\end{figure}

\begin{figure}[h]
 \begin{subfigure}[h]{.5\textwidth}
   \centering
  \includegraphics[width=.85\linewidth]{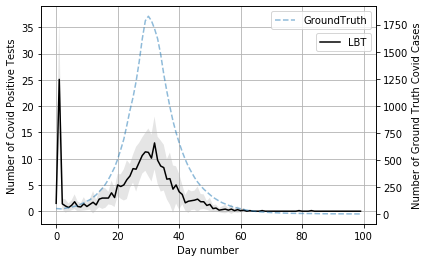}
    \caption{Daily number of tests.}
\end{subfigure}
 \begin{subfigure}[h]{.5\textwidth}
      \centering
  \includegraphics[width=.85\linewidth]{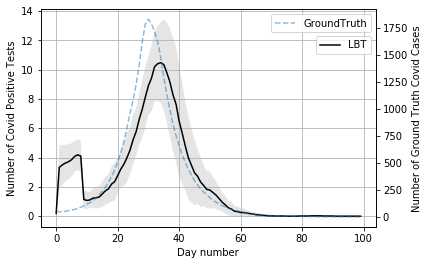}
  \caption{Daily number of tests smoothened by taking an average over a window of 8 days.}
\end{subfigure}
\caption{Evolution of COVID-19 cases under the indefinite lockdown intervention using the thresholded smoothened slope policy.}
\label{f:lockdown_LBT}
\end{figure}

We have implemented a threshold-based lockdown policy which disables
the interaction of the agents with each other once the slope of the
number of positive tests graph crosses a threshold. Specifically,
denoting by $P(t)$ the number of positive test outcomes, our policy
computes the slope of the ``10 day chord'' of a smoothened version $\overline{P}(t):=\frac 1 8\sum_{i=0}^7P(t-i)$ of $P(t)$ given by
\[
\theta(t)=\frac{\overline{P}(t)-\overline{P}(t-10)}{10};
\]
and starts a lockdown once $\theta(t)$ crosses a fixed threshold $\tau=0.5$. The selection of slope of the smoothened graph of $P(t)$ as a feature to use is based on our empirical observation that this slope captures the slope of the ground truth COVID-19 cases graph well.
We term this policy the {\em thresholded smoothened slope} policy for lockdown.
We use the same clustered seed as in the previous section and 
present our results using RST, CT, and LBT in
Figures~\ref{f:lockdown_RST},~\ref{f:lockdown_CT},
and~\ref{f:lockdown_LBT}, respectively.

Looking at the ground truth number of cases under three testing policies, we observe that CT reduces the maximum daily COVID-19 cases the most by about 90\% of the peak value in absence of any intervention. Even RST offers a significant reduction, by about 80\%, but is much worse than CT. Remarkably, LBT, too offers very similar reduction as CT while requiring much less operational effort. 
\subsection{Fixed duration lockdown}
\begin{figure}[h]
 \begin{subfigure}[h]{.5\textwidth}
   \centering
  \includegraphics[width=.85\linewidth]{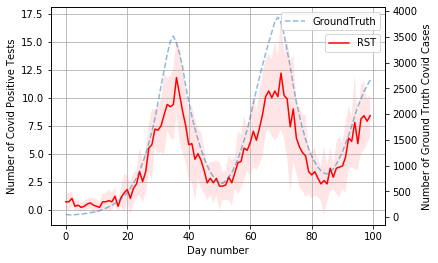}
    \caption{Daily number of tests.}
\end{subfigure}
 \begin{subfigure}[h]{.5\textwidth}
      \centering
  \includegraphics[width=.85\linewidth]{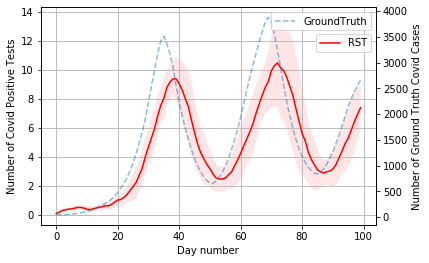}
  \caption{Daily number of tests smoothened by taking an average over a window of 8 days.}
\end{subfigure}
\caption{Evolution of COVID-19 cases under the fixed duration lockdown intervention using the thresholded smoothened slope policy.}
\label{f:lockdownfixed_RST}
\end{figure}

\begin{figure}[h]
 \begin{subfigure}[h]{.5\textwidth}
   \centering
  \includegraphics[width=.85\linewidth]{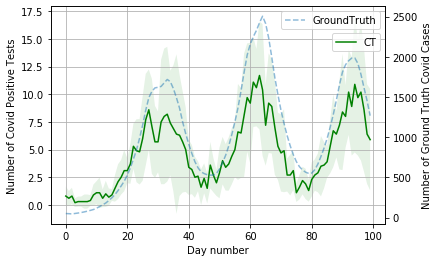}
    \caption{Daily number of tests.}
\end{subfigure}
 \begin{subfigure}[h]{.5\textwidth}
      \centering
  \includegraphics[width=.85\linewidth]{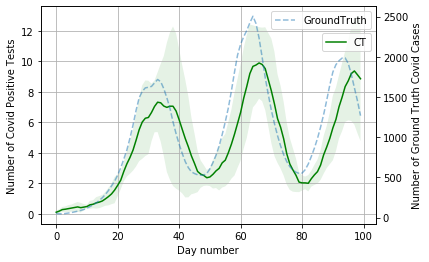}
  \caption{Daily number of tests smoothened by taking an average over a window of 8 days.}
\end{subfigure}
\caption{Evolution of COVID-19 cases under the fixed duration lockdown intervention using the thresholded smoothened slope policy.}
\label{f:lockdownfixed_CT}
\end{figure}

\begin{figure}[h]
 \begin{subfigure}[h]{.5\textwidth}
   \centering
  \includegraphics[width=.85\linewidth]{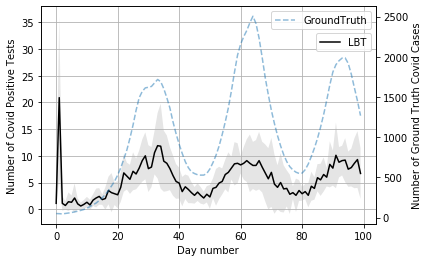}
    \caption{Daily number of tests.}
\end{subfigure}
 \begin{subfigure}[h]{.5\textwidth}
      \centering
  \includegraphics[width=.85\linewidth]{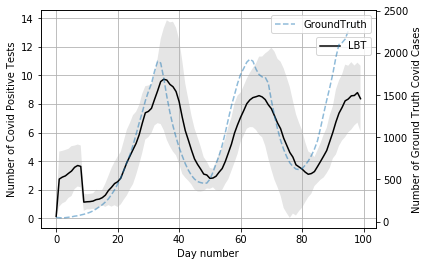}
  \caption{Daily number of tests smoothened by taking an average over a window of 8 days.}
\end{subfigure}
\caption{Evolution of COVID-19 cases under the fixed duration lockdown intervention using the thresholded smoothened slope policy.}
\label{f:lockdownfixed_LBT}
\end{figure}

The last intervention we consider is a fixed duration lockdown which, too, uses the thresholded smoothened slope policy to engage a lockdown. But after a lockdown is initiated, it is lifted after 14 days. 
We use the same clustered seed as in the previous sections and 
present our results using RST, CT, and LBT in
Figures~\ref{f:lockdownfixed_RST},~\ref{f:lockdownfixed_CT},
and~\ref{f:lockdownfixed_LBT}, respectively.

We observe that each testing policy results in three such lockdowns over a hundred day period in our simulation. Interestingly, even with significantly fewer days of lockdown, the reduction in the number of ground truth COVID-19 cases is only marginally less than indefinite lockdown of the previous section. Further, here, too,CT outperforms RST significantly and LBT yields similar performance as CT.  

\section{Concluding Remarks}\label{s:conclusions}
\begin{figure}[h]
 \begin{subfigure}[h]{.5\textwidth}
   \centering
  \includegraphics[width=.85\linewidth]{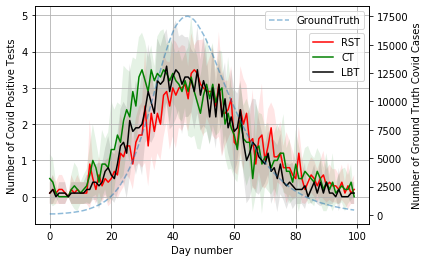}
    \caption{Daily number of tests.}
\end{subfigure}
 \begin{subfigure}[h]{.5\textwidth}
      \centering
  \includegraphics[width=.85\linewidth]{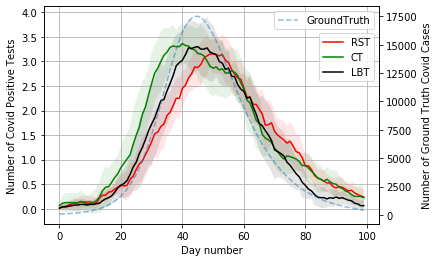}
  \caption{Daily number of tests smoothened by taking an average over a window of 8 days.}
\end{subfigure}
\caption{Comparative test performance with uniform seeding and without intervention, for a time period of 100 days and with a {\bf testing budget of 5 tests/day}. Results are averaged across 10 runs and error bars represent 1 standard deviation.}
\label{f:5tests}
\end{figure}

\begin{figure}[h]
 \begin{subfigure}[h]{.5\textwidth}
   \centering
  \includegraphics[width=.85\linewidth]{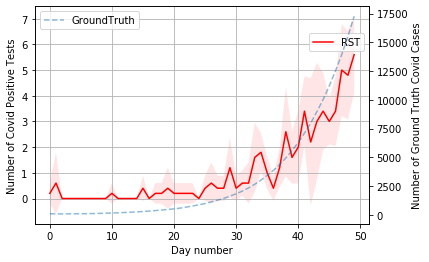}
    \caption{Daily number of tests.}
\end{subfigure}
 \begin{subfigure}[h]{.5\textwidth}
      \centering
  \includegraphics[width=.85\linewidth]{RST_USeed_1M.png}
  \caption{Daily number of tests smoothened by taking an average over a window of 8 days.}
\end{subfigure}
\caption{Performance of RST with uniform seeding and without intervention for a {\bf population of 1 Million}, for a time period of 50 days, and with a {testing budget of 50 tests/day}. Results are averaged across 5 runs and error bars represent 1 standard deviation.}
\label{f:1Mpop}
\end{figure}

We have taken an initial stab at the question we set to answer: How
well do test results reveal the ground truth and help decide  when to  interventions? Using our simulation
framework with 1 test per 2000 people, we have observed that the
simplest testing strategy RST captures the trend well. Other
strategies such as CT and our proposed LBT yield higher number of
positive tests than RST, 
and are seen to be more effective in enabling interventions. Note that, at the time of writing, 
this number of daily tests is much higher than what is being done in
India. 
Roughly, 30,000 tests are done daily in India when this article is
being written and roughly 1,500 of them happen in the state of Karnataka, the state to
which the city of Bengaluru belongs. This amounts to 1 test for about
40,000 people. In fact, even at this low number of tests,  positive tests
reflect the ground truth trend in our experiments. 
We conducted experiments with uniform seeding with 5 tests per day
(for 100K population) and a shorter experiment with RST alone with 50
tests per day for 1M population; the results are displayed in
Figures~\ref{f:5tests} and~\ref{f:1Mpop}.  

We are tempted to conclude from these observations that India is doing
enough daily tests to get a good idea of the ground truth trends. 
Even with these few tests, a mix of RST and CT, which seems to be the
current testing policy for India and many other countries, is a good
testing policy. 
In the initial phase of the spread, perhaps one can control the rapid
spread by identifying a large fraction of the infected patients by
increasing testing to scale up to the number of infected. But once a
significant fraction of population is infected, it may become
infeasible to quarantine all the patients. In such a situation, non
pharmaceutical interventions such as lockdown are the only option and
the role of testing is to get a handle on the ground truth trend. Our
simulations suggest that this can be done even with 1 test per 10000
people or so.   

But there are a few cautionary remarks we must make: First, the
sampling probabilities that emerge during test prescription by doctors
in practice are difficult to evaluate and may deviate significantly
form our ideal assumptions. Second, our policies rely on symptomatic
patients coming out for testing (CT does not assume that). While this
can be ensured by an active information campaign and ready access to
doctors and digital medical advice, it is important to understand what
can happen when a subsection of population is unable to report
symptoms. Third, having more positive tests can perhaps offer
robustness to  deviations from our model in practice. 

Finally, we remark that there were many omissions from our initial
study. We have not studied gradual ramping-up of test budget, nor have
we studied more localized interventions (though our simulations
suggest that LBT will be more effective to enable a local
lockdown). The effect of under reporting needs to be studied more
thoroughly, too. It is of interest also to study extensive testing in
an emerging hot-spot as well as pooling for more efficient testing. We
plan to consider these issues in later versions of this ongoing work. 

\section*{Acknowledgements} We thank Rajesh Sundaresan (IISc) for useful discussions and encouragement to build a simulation framework for testing COVID-19,
Abdul Pinjari (IISc) for sharing Bengaluru traffic flow data available at CiSTUP, IISc, and Ph.D. students Debangshu Banerjee,
Sayak Ray Chowdhury, Lekshmi Ramesh, and K. R. Sahasranand for help with data handling. 
\bibliographystyle{IEEEtranS}
\bibliography{references}

\end{document}